\documentclass[aip,amsmath,amssymb,reprint]{revtex4-1}

\usepackage{graphicx}% Include figure files
\usepackage{dcolumn}% Align table columns on decimal point
\usepackage{bm}% bold math
\usepackage[utf8]{inputenc}
\usepackage[T1]{fontenc}
\usepackage{mathptmx}
\usepackage{etoolbox}
\usepackage{float}
\usepackage[normalem]{ulem}

\newcommand{\reffig}[1]{FIG. \ref{#1}}
\newcommand{\refeqn}[1]{equation (\ref{#1})}
\newcommand{\refp}[1]{(\ref{#1})}

\makeatletter
\def\@email#1#2{%
 \endgroup
 \patchcmd{\titleblock@produce}
  {\frontmatter@RRAPformat}
  {\frontmatter@RRAPformat{\produce@RRAP{*#1\href{mailto:#2}{#2}}}\frontmatter@RRAPformat}
  {}{}
}%
\makeatother

\begin{document}

\preprint{AIP/123-QED}

\title{Brachistochrone of off-centered cylinders}

\author{Krishnaraj Sambath}
\email{ksambath@chevron.com}
\affiliation{Chevron Corporation, Midland, TX, USA}
\author{Vidhya Nagarajan}
\affiliation{ChampionX, Sugarland, TX, USA}

\date{19 December 2022}

\begin{abstract}
We consider the problem of finding paths of shortest transit time between two points (popularly known as Brachistochrone) for cylinders with off-centered center of mass, rolling down without slip,  subject solely to the force of gravity. This problem is set up using principles of classical rigid body dynamics and the desired path function is solved for numerically using the method of discrete calculus of variations. We discover a distinct array of brachistochrone trajectories for off-centered cylinders, demonstrate a critical dependence of such paths on the initial location and orientation of cylinders' centers of mass and bring new insights into the family of brachistochrone problems and solutions.
\end{abstract}

\maketitle

\section{Introduction}

\begin{figure}
    \centering
    \includegraphics[width=0.4\textwidth]{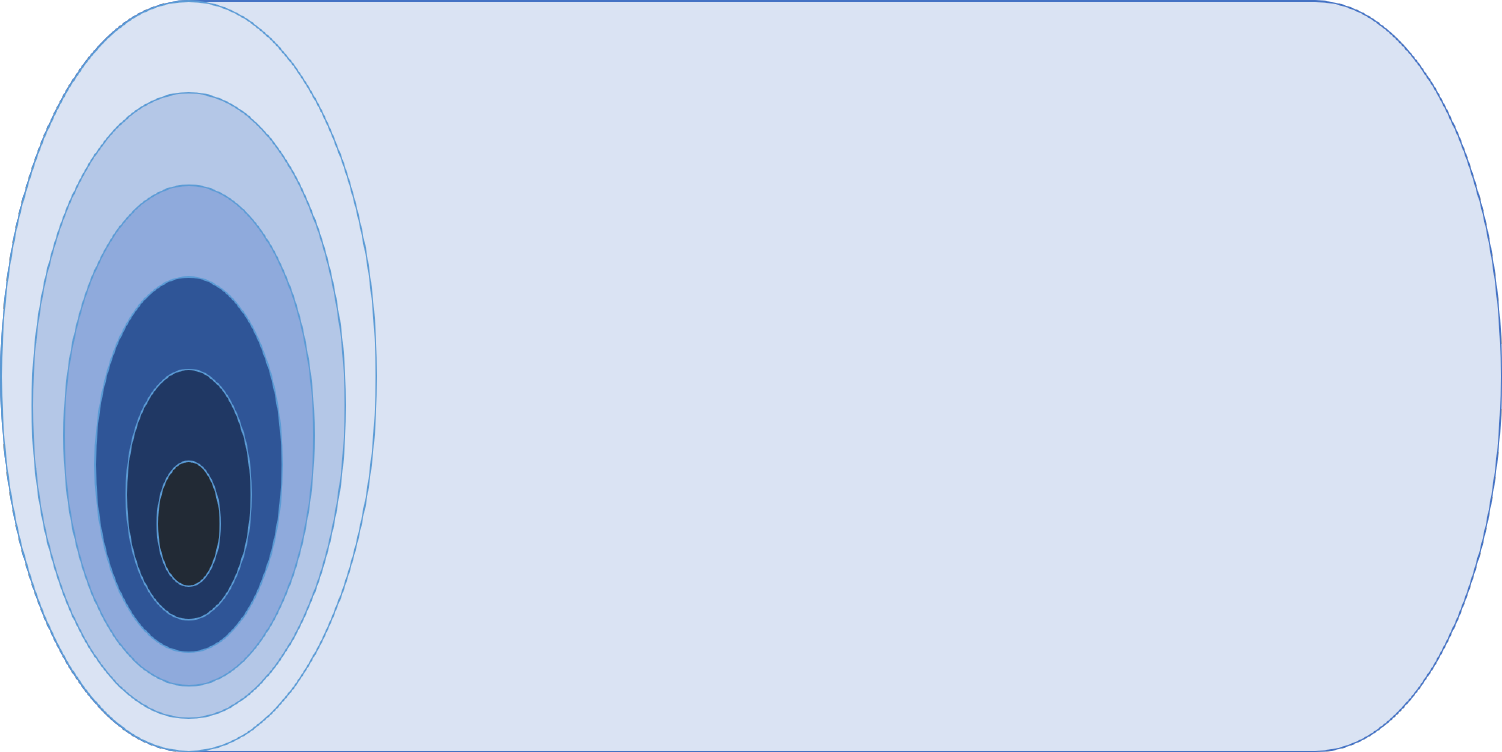}
    \caption{Example of non-homogeneous (off-centered) rigid cylinder, where darker shades represent denser sections}
    \label{figSketch}
\end{figure}

The brachistochrone problem concerns the path of shortest transit time between two points for an object traversing solely under the force of gravity. This problem has a celebrated history, having attracted the best minds in the field of mathematics and physics for over three centuries \citep{babb2012brachistochrone}. The solution to that problem for a simple point mass (bead) is a cycloid, which has been deduced by a variety of methods (Calculus of variations\cite{benson1969elementary}, Geometry\cite{herrera1994galileo}, Laws of refraction\cite{erlichson1999johann}, Optimal Control Theory\cite{sussmann2002brachistochrone}). Several variations and generalizations of the problem has been proposed (with different force fields, types of object traversing the path) and solved \citep{rodgers1946brachistochrone,ashby1975brachistochrone,goldstein1986relativistic,parnovsky1998some,akulenko2009brachistochrone,legeza2010brachistochrone,obradovic2014brachistochronic,stevens2017brachistochrone,gurram2019brachistochrone,benham2020brachistochrone}. In all of the previous works, objects considered were beads, cylinders or other axisymmetric objects. In this work, we extend the literature of Brachistochrone to off-centered objects whose centers of mass (\textit{c.o.m}) are not necessarily coincident with their centers of cylinder (\textit{c.o.c}). Such objects maybe realized with non-uniform spatial distribution of mass, an example of which is illustrated in \reffig{figSketch}.

\section{Problem Formulation} \label{sPF}

\subsection{Governing Equations}

Consider a cylinder of radius $r$, mass $m$, with its \textit{c.o.m} located $r_m = \eta_m r$ (with $0\le \eta_m\le1$) away from its \textit{c.o.c} and moment of inertia of $I_{cm} = kmr^2$ about an axis through its \textit{c.o.m} and parallel to axis of geometric symmetry. The mass distribution within the cylinder determines the value of $k$, whose range would be $0\le k\le 1$. For example, a hollow cylinder, with all its mass distributed uniformly on the outer edge would have a value of $k = 1$ whereas a dense rod at the center connected to massless but rigid cylindrical shell with massless spokes would have a $k = 0$. Parameters $k$ and $\eta_m$ are not entirely independent as they both depend on the mass distribution within the circular frame of the cylinder. Considering the extremities of density distribution (for a given $m$ and $r$), it is deduced that for a given $\eta_m$, the maximum value $k$ can take is $1 - \eta_m^2$. Some example configurations of off-centered cylinders are shown in Appendix A.

\begin{figure}
    \centering
    \includegraphics[width=0.48\textwidth]{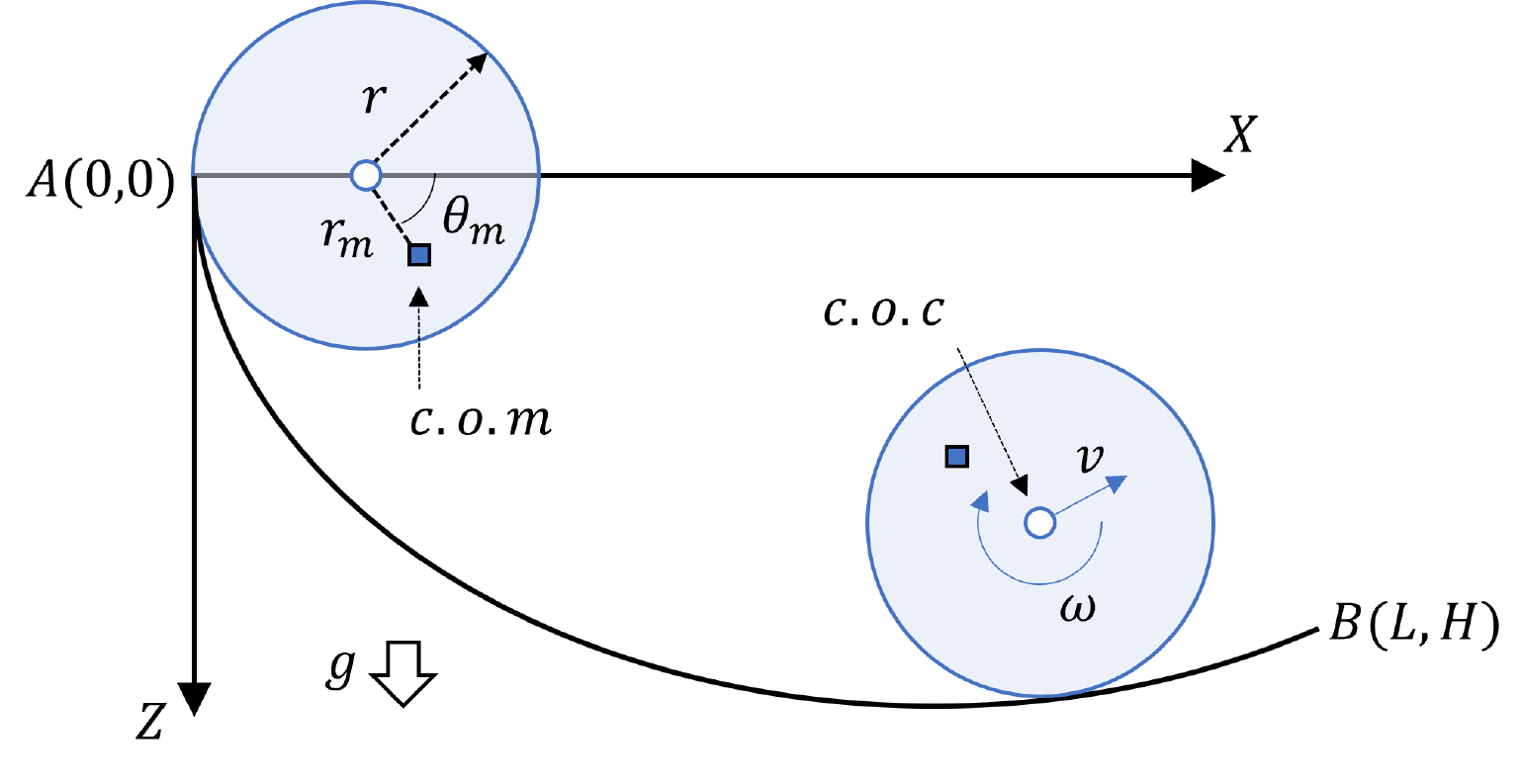}
    \caption{Problem sketch}
    \label{figSketch2}
\end{figure}

Let the two points between which the cylinder would be rolling be defined as A (0, 0) and B ($L$, $H$) in cartesian coordinates system of $(X,Z)$, with direction of gravity oriented along the $+Z$ axis. Let the trajectory of the \textit{c.o.c} of cylinder, as it rolls down from A to B, be described by $z_c(x_c)$. The desired brachistochrone curve (path on which the cylinder rolls) is related to, but different from, $z_c(x_c)$. This curve will be defined at the end of this section. Let the initial configuration of cylinder be such that its \textit{c.o.m} makes an angle of $\theta_m$ with the horizontal axis as shown in \reffig{figSketch2}.

Conservation of energy between potential and kinetic for such a system is given as follows\cite{carnevali2005rolling,hu2011rolling,hojgaard2012five}.

\begin{equation}
    \frac{1}{2}mv_{cm}^2+\frac{1}{2}I_{cm}\omega^2=mg (z_c+r_m (\sin \theta - \sin \theta_m)) \label{eqnEnergyBalance}
\end{equation}

Here, $v_{cm}$ represents the linear speed of the \textit{c.o.m}, $\omega$ the angular speed, $\theta$ the angular displacement of \textit{c.o.m} and $g$ acceleration due to gravity. Note that \refeqn{eqnEnergyBalance} is independent of the axial location of the \textit{c.o.m} of the object ($y_m$ - in the direction perpendicular to the plane of sketch in \reffig{figSketch2}) and therefore the results obtained here hold true for all off-centered cylinders with a given ($r_m$, $\theta_m$) regardless of $y_m$.

Location of cylinder's \textit{c.o.m} is related to that of its \textit{c.o.c} as follows

\begin{align}
    x_{cm} & = x_c + r_m \cos \theta \label{eqnxcm} \\
    z_{cm} & = z_c + r_m \sin \theta \label{eqnzcm}
\end{align}

Time derivatives (denoted by $\cdot$ on variables) of above equations yield the below relationships for speeds of cylinder's \textit{c.o.m} and \textit{c.o.c}.

\begin{align}
    v_{cm} & = \sqrt{\dot{x}_{cm}^2+\dot{z}_{cm}^2} \label{eqnvcm} \\
    v_c & = \sqrt{\dot{x}_c^2+\dot{z}_c^2} \label{eqnv}
\end{align}

No-slip condition between the cylinder and path establishes the following relationship\cite{rodgers1946brachistochrone} between $v_c$ and $\omega$.

\begin{equation}
    v_c = r\omega = r\dot{\theta} \label{eqnNoslip}
\end{equation}

Combining equations \refp{eqnEnergyBalance} - \refp{eqnNoslip}, time taken for the cylinder to roll down from A to B is obtained as follows (details of derivation are given in Appendix B).

\begin{equation}
    T[z_c(x_c)] = \int_0^{L} \frac{\sqrt{1 + \eta_m^2 - 2 \eta_m \sin(\theta - \beta) + k}}{\sqrt{2g(z_c+r_m (\sin \theta - \sin \theta_m))}}\sqrt{1 + z_c'^2}dx_c \label{eqnTransitTime}
\end{equation}

Here, $z_c'$ indicate the derivatives of $z_c(x_c)$ with respect to $x_c$ and $\beta$ is related to the instantaneous slope of the curve as follows.

\begin{equation}
    \beta = \tan^{-1}(z_c') \label{eqnbeta}
\end{equation}

The no-slip condition is re-written in terms $z_c(x_c)$ and $\theta(x_c)$ to obtain the following relation.

\begin{equation}
    \theta(x_c) = \theta_m + \frac{1}{r}\int_0^{x_c} \sqrt{1 + z_c'^2} dx_c \label{eqnThetaInt}
\end{equation}

The desired path of least time on which the cylinder rolls down $z_c(x_c)$ is obtained by minimizing the functional $T[z_c(x_c)]$.

\begin{equation}
    \frac{\delta T[z_c(x_c)]}{\delta z_c(x_c)} = 0 \label{eqnTimeFuncDeriv}
\end{equation}

With trajectory of \textit{c.o.c} solved for, the desired brachistochrone path (on which the cylinder rolls) is then obtained by tracking the point at a distance $r$ away and normal to $z_c(x_c)$.  Such a path, defined as $z_b(x_b)$, is given by the following equations.

\begin{align}
    x_b & = x_c - r \sin\beta \label{eqnxb} \\
    z_b & = z_c + r \cos\beta \label{eqnzb}
\end{align}

From \refeqn{eqnTransitTime}, it can be inferred that the brachistochrone not only depends on the location of \textit{c.o.m} ($\eta_m$ and $\theta_m$) but also on the moment of inertia factor ($k$). This is in contrast with axisymmetric objects where their brachistochrone paths are independent of their moments of inertia\cite{rodgers1946brachistochrone}. It can also be verified that, for the case of $\eta_m = 0$, the above equations simplify to the familiar system for a uniform cylinder\cite{babb2012brachistochrone}.

\subsection{Solution Methodology}

The manner of coupling between the equations \refp{eqnEnergyBalance} - \refp{eqnzb} does not lend them to analytical solutions using classical calculus of variations (barring the limiting case of $\eta_m = 0$). Therefore, a discretized and numerical approach\cite{cadzow1970discrete,tan1988discrete} is taken here. Additionally, steep gradients in the solution for $z_c(x_c)$ prohibit a direct numeric solution. Therefore, the system is recast in parametric coordinates as $x_c(\xi), z_c(\xi)$, where $\xi$ has a range of $0\le\xi\le1$ and represents dimensionless path length along the curve $z_c(x_c)$, as measured from A. This parameterized curve is discretized into $n$ points $\{x_c(\xi_j)=x_j, z_c(\xi_j)=z_j; j = 1, 2, ..., n\}$. These points along the curve are equally spaced by defining $\xi_j := (j - 1) / (n - 1)$. This method is akin to arc length formulation commonly used to solve moving boundary problems with complex domains in fluid mechanics\cite{sambath2014electrohydrostatics,sambath2019inertial}.

Governing equations for $x_j$'s are given by equations \refp{eqnx1} - \refp{eqnxn}, where, equations \refp{eqnx1} and \refp{eqnxn} enforce the boundary conditions and equation set \refp{eqnxj} solve for the $x$-grid points.

\begin{align}
    x_1 - \frac{rz_1'}{\sqrt{1+z_1'^2}} & = 0 \hspace{10pt} \text{for} \hspace{5pt} j = 1 \label{eqnx1}\\
    \sqrt{(x_j - x_{j-1})^2 + (z_j - z_{j-1})^2} - S \Delta \xi & = 0 \hspace{10pt} \text{for} \hspace{5pt} j = 2, 3, ... n \label{eqnxj}\\
    x_n - \frac{r z_n'}{\sqrt{1+z_n'^2}} & = L \hspace{9pt} \text{for} \hspace{5pt} S\label{eqnxn}
\end{align}

Here, $S$ represents the total arc length of the curve, which is defined and computed by \refeqn{eqnS}.

\begin{equation}
    S = \sum_{j=2}^n \sqrt{(x_j-x_{j-1})^2 + (z_j-z_{j-1})^2} \label{eqnS}
\end{equation}

Governing equations for $z_j$'s are given by equations \refp{eqnz1} - \refp{eqnzn}, where equations \refp{eqnz1} and \refp{eqnzn} enforce the boundary conditions and equation set \refp{eqnzj} correspond to the discretized version of functional minimization, \refeqn{eqnTimeFuncDeriv}.

\begin{align}
    z_1 + \frac{r}{\sqrt{1+z_1'^2}} & = 0 \hspace{10pt} \text{for} \hspace{5pt} j = 1 \label{eqnz1} \\
    \frac{\partial T}{\partial z_j} \approx \frac{T(z_j + \varepsilon) - T(z_j - \varepsilon)}{2\varepsilon} & = 0 \hspace{10pt} \text{for} \hspace{5pt} j = 2, 3, ..., n - 1 \label{eqnzj} \\
    z_n + \frac{r}{\sqrt{1+z_n'^2}} & = H \hspace{10pt} \text{for} \hspace{5pt} j = n \label{eqnzn}
\end{align}

For derivatives of $z_c(x_c)$ at the boundaries, first order forward or backward differences method, as appropriate, is applied. In equation set \refp{eqnzj}, $\varepsilon$ is chosen to be small number ($10^{-3}$) to compute the approximated partial derivative of $T(z_j)$ with respect to $z_j$.

\subsection{Benchmarking and Validation Tests}

\begin{figure}
     \centering
    \includegraphics[width=0.48\textwidth]{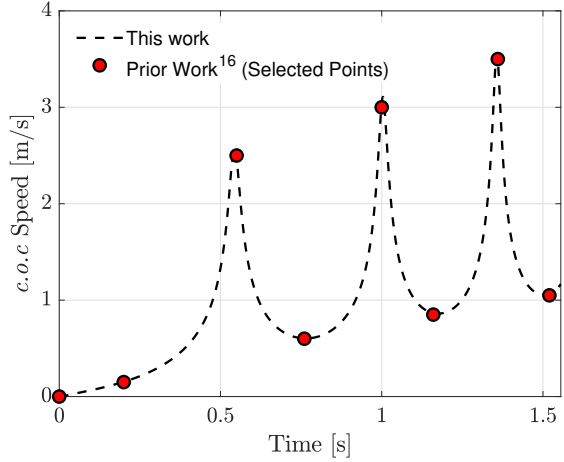}
    \caption{Time evolution of instantaneous speed of \textit{c.o.m} of an off-centered cylinder rolling down an inclined plane, compared with results of prior studies\cite{carnevali2005rolling} for $m = 1.195$ $kg$, $r = 0.077$ $m$, $k = 0.255$, $\eta_m=0.6521$, $\theta_m=-83.25^0$, $L = 1.49$ $m$, $H = 0.18$ $m$}
    \label{figVerification}
\end{figure}

Equations \refp{eqnx1} - \refp{eqnzn} form a system of coupled non-linear equations with $2n+1$ variables, which are solved using MATLAB \texttt{fsolve} function. Four independent tests were performed to evaluate the validity of the equations developed, numerical approach employed and code written:

\begin{itemize}
    \item Code developed is used to simulate rolling dynamics of off-centered cylinder down an inclined plane (by forcing the curve between the end points to be a straight line) for previously studied systems. \reffig{figVerification} shows the transients of instantaneous speed of cylinder's \textit{c.o.c} and how it matches the results reported in prior studies\cite{carnevali2005rolling}
    \item Code developed is used to determine the brachistochrone for uniform cylinder (by setting $\eta_m=0$) traversing from A(0, 0) to B ($r + \pi$, $r + 2$). The shortest transit time was computed to be 1.229 $s$, which matches the analytical solution of $\pi\sqrt{3/2g}$ reported in prior work\cite{legeza2010brachistochrone} within a tolerance of 0.05\%
    \item The brachistochrone results for off-centered cylinders ($L = 3$, $r = 1/3$, $\eta_m = 2/3$, $\theta_m = 0$, $k = 0.5$) are verified to be grid independent by comparing the paths $z_c(x_c)$ for various values of $n$. Additionally, the dimensionless time of transit, $\tau := T[z_c(x_c)] \sqrt{2g/H}$, corresponding to $n = 25, 50, 100$ and $200$ are seen to converge as 4.97, 4.99, 4.99 and 4.99 respectively. For the remainder of the solutions presented here, $n$ is set to 100
    \item Accuracy of numerical solutions are ensured by keeping the absolute value of norm of residual errors and solution updates below $10^{-6}$ for all results reported here
\end{itemize}

Without loss of generality, the vertical distance ($H$) between points A and B is set to 1 length unit, making it the reference length scale for this problem. For each of the parameter values studied, the system of equations (\ref{eqnx1}) - (\ref{eqnzn}) is solved. The brachistochrone path $z_b(x_b)$ and corresponding shortest transit time are then obtained using equations (\ref{eqnxb}) - (\ref{eqnzb}) and (\ref{eqnTransitTime}), respectively.

\section{Results}

\begin{figure}
     \centering
    \includegraphics[width=0.48\textwidth]{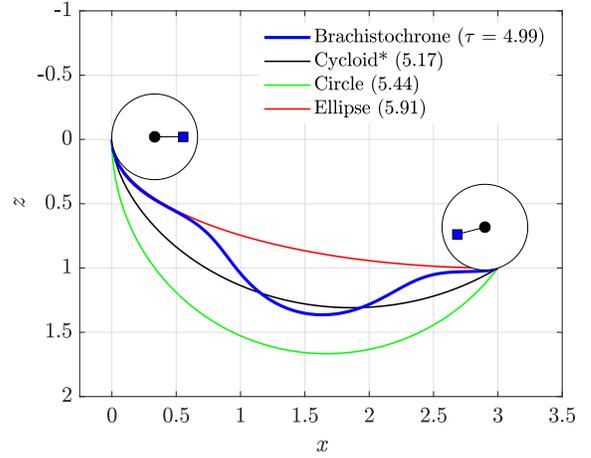}
    \caption{Brachistochrone (path of shortest transit time) of an off-centered cylinder ($r = 1/3$, $\eta_m=2/3$, $\theta_m = 0$, $k = 0.5$) for a domain size of $L=3$. For comparison, few other standard curves and corresponding transit times for this cylinder over the same domain, are shown}
    \label{figDemoCasePaths}
\end{figure}

\reffig{figDemoCasePaths} shows the brachistochrone path for a representative off-centered cylinder ($r = 1/3$, $\eta_m=2/3$, $\theta_m = 0$, $k = 0.5$) over a domain size of $L = 3$. The transit time for the cylinder through this path is 4.99. For comparison, few other standard curves between the same two points and corresponding transit times for the same cylinder are shown. The path marked as `Cycloid*' indicates a path generated by the outer surface of the cylinder as its \textit{c.o.c} traverses a cycloid (Such a path corresponds to brachistochrone of an uniform cylinder of same radius $r$).

\subsection{Feasibility of brachistochrone path}

\begin{figure}
     \centering
    \includegraphics[width=0.48\textwidth]{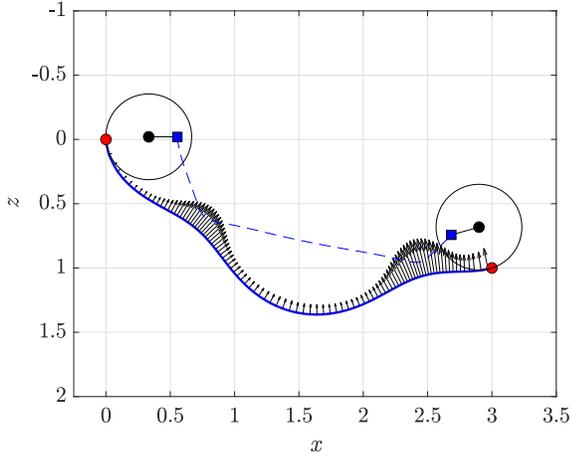}
    \caption{Normal reaction force ($R_n$) on the cylinder as it rolls along its brachistochrone path (system parameters same as that in \reffig{figDemoCasePaths}). $R_n$ reaches a maximum value of $4.2mg$ at $x=2.5$}
    \label{figDemoCaseForces}
\end{figure}

Two conditions of feasibility of the obtained brachistochrone path are evaluated: (a) normal reaction force on the cylinder and (b) curvature of the path.

The normal reaction force on the cylinder, $R_n = \textbf{R}.\textbf{n}$ is deduced from the force balance on the cylinder\cite{legeza2010brachistochrone} given by \refeqn{eqnForce}. The acceleration of \textit{c.o.m} is given by \refeqn{eqnacm} which is obtained by taking time derivatives of equations (\ref{eqnxcm}) - (\ref{eqnzcm}), twice. Time taken for cylinder to reach any given point along the path is obtained using \refeqn{eqnTransitTime}. Unit vector normal to the path is given by \refeqn{eqnnormal}. In these equations, $\textbf{e}_x$ and $\textbf{e}_z$ represent unit vectors in $X$ and $Z$ directions, respectively. \reffig{figDemoCaseForces} shows the normal reaction force remains positive ($R_n > 0$) along the entire path confirming that the cylinder will remain in contact with its path as it rolls down. The normal force increases as the \textit{c.o.m} rolls towards the path and decreases as \textit{c.o.m} rolls away. Quantitatively, the maximum normal reaction force experienced, for this case, is $4.2mg$, with bulk of it used to counter the centripetal force generated by the \textit{c.o.m} rotating around \textit{c.o.c}.

\begin{equation}
    m\textbf{a}_{cm} = mg\textbf{e}_z + \textbf{R} \label{eqnForce}
\end{equation}

\begin{equation}
    \textbf{a}_{cm} = \ddot{x}_{cm}\textbf{e}_x + \ddot{z}_{cm}\textbf{e}_z \label{eqnacm}
\end{equation}

\begin{equation}
    \textbf{n} = \frac{\dot{z}_b \textbf{e}_x - \dot{x}_b\textbf{e}_z}{(\dot{x}_b^2+\dot{z}_b^2)^{1/2}} \label{eqnnormal}
\end{equation}

As a second feasibility check, the curvature of the path $z_b(x_b)$, is computed using \refeqn{eqnkappa} and compared against that of the cylinder ($\kappa_0=1/r$). For the brachistochrone curve in \reffig{figDemoCasePaths}, $\kappa_b$ varied between (-1.34, 1.52) staying well below $\kappa_0 = 3$. This confirms that this brachistochrone path does not have sharp enough turns to physically prevent the cylinder from rolling on those turns.

\begin{equation}
    \kappa_b = \frac{\dot{x}_b\ddot{z}_b-\ddot{x}_b\dot{z}_b} {(\dot{x}_b^2 + \dot{z}_b^2)^{3/2}} \label{eqnkappa}
\end{equation}

For all of the cases studied in the following subsections, the normal reaction forces were verified to remain positive ($R_n > 0$) throughout the cylinders' motions on their respective brachistochrone paths and the curvatures of those paths were verified to be less than that of the cylinder ($\kappa_b < \kappa_0$), confirming the feasibility of the solutions presented.

\subsection{Comparison against other objects}

\begin{figure}
     \centering
    \includegraphics[width=0.48\textwidth]{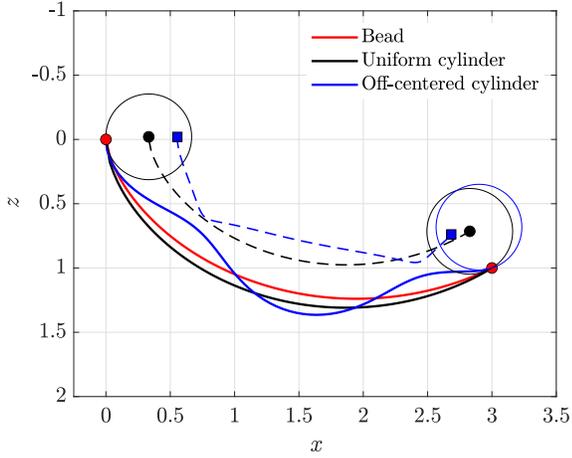}
    \caption{Brachistochrones of bead, uniform cylinder and off-centered cylinder ($r = 1/3$, $\eta_m=2/3$, $\theta_m = 0$, $k = 0.5$) for a domain size of $L=3$}
    \label{figDemoCase}
\end{figure}

\reffig{figDemoCase} shows brachistochrone paths (indicated by solid lines) for a bead, uniform cylinder and off-centered cylinder for a representative set of system parameters. Shape of bead's brachistochrone is a cycloid\cite{herrera1994galileo}. Shape of uniform cylinder's brachistochrone is a curve traced by its outer edge as its \textit{c.o.m} traverses a cycloid\cite{legeza2010brachistochrone}. The off-centered cylinder's brachistochrone presents a richer curve with more features, referred to here as "valleys and peaks". Dashed lines in \reffig{figDemoCase} indicate trajectories of \textit{c.o.m} of the two cylinders - uniform (black) and off-centered (blue). As the off-centered cylinder is rolling from A to B, the brachistochrone path is "pulled-in" when the \textit{c.o.m} is at the bottom and "pushed-out" when the \textit{c.o.m} is at the top. Such a path facilitates the \textit{c.o.m} to not fall too deep and lose too much potential energy which it will then need to gain back to reach B. In terms of path lengths, bead has the shortest path (of 3.57) among the three, with uniform cylinder and off-centered cylinder having near equal path lengths (of 3.66). In terms of transit times, the bead is the quickest ($\tau = $ 4.52). However, the off-centered cylinder comes second (4.99) ahead of uniform cylinder (5.12), despite having near equal path lengths. Similar trends (of off-centered cylinder having quicker transit times compared to that of uniform cylinder) have been reported for standard inclined plane\cite{carnevali2005rolling}.

\begin{figure}
        \centering
        \includegraphics[width=0.48\textwidth]{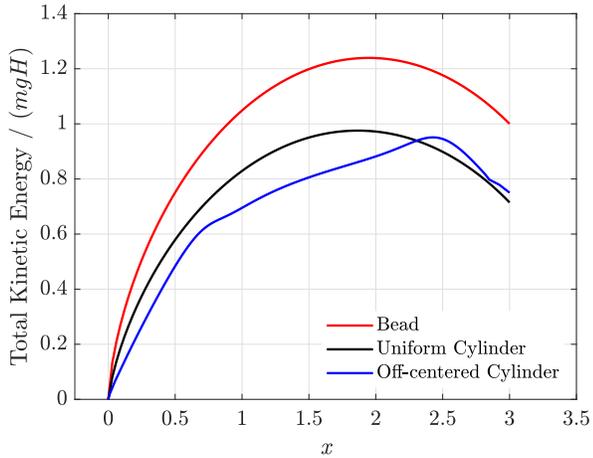}
        \caption{Evolution of total kinetic energy (non-dimensionalized with $mgH$) for bead, uniform cylinder and off-centered cylinder (parameters as in \reffig{figDemoCase}) as they traverse through their brachistochrones}
        \label{figTKEx}
\end{figure}

Compared to beads, uniform cylinder is known to take a longer time due to the additional inertia from its rotational motion\cite{rodgers1946brachistochrone}. In the case of off-centered cylinders, this exchange of energy between kinetic and gravitational forms is influenced by the initial location of \textit{c.o.m} and its subsequent trajectory. \reffig{figTKEx} compares the kinetic energy evolution for a bead, uniform cylinder and off-centered cylinder. The trends of instantaneous linear and angular speeds of the cylinders will follow that of the total kinetic energy as they are directly related by $\omega \propto \sqrt{TKE}$ and \refeqn{eqnNoslip}.

In the following subsections, the effects of various parameters - initial locations of \textit{c.o.m} in the angular ($\theta_m$) and radial ($\eta_m$) directions, moment of inertia ($k$), radius ($r$) of cylinder and length of domain ($L$) - on the brachistochrone paths and corresponding shortest transit times are systematically studied.

\subsection{Effect of radial location of \textit{c.o.m}}

\begin{figure}
    \centering
    \includegraphics[width=0.48\textwidth]{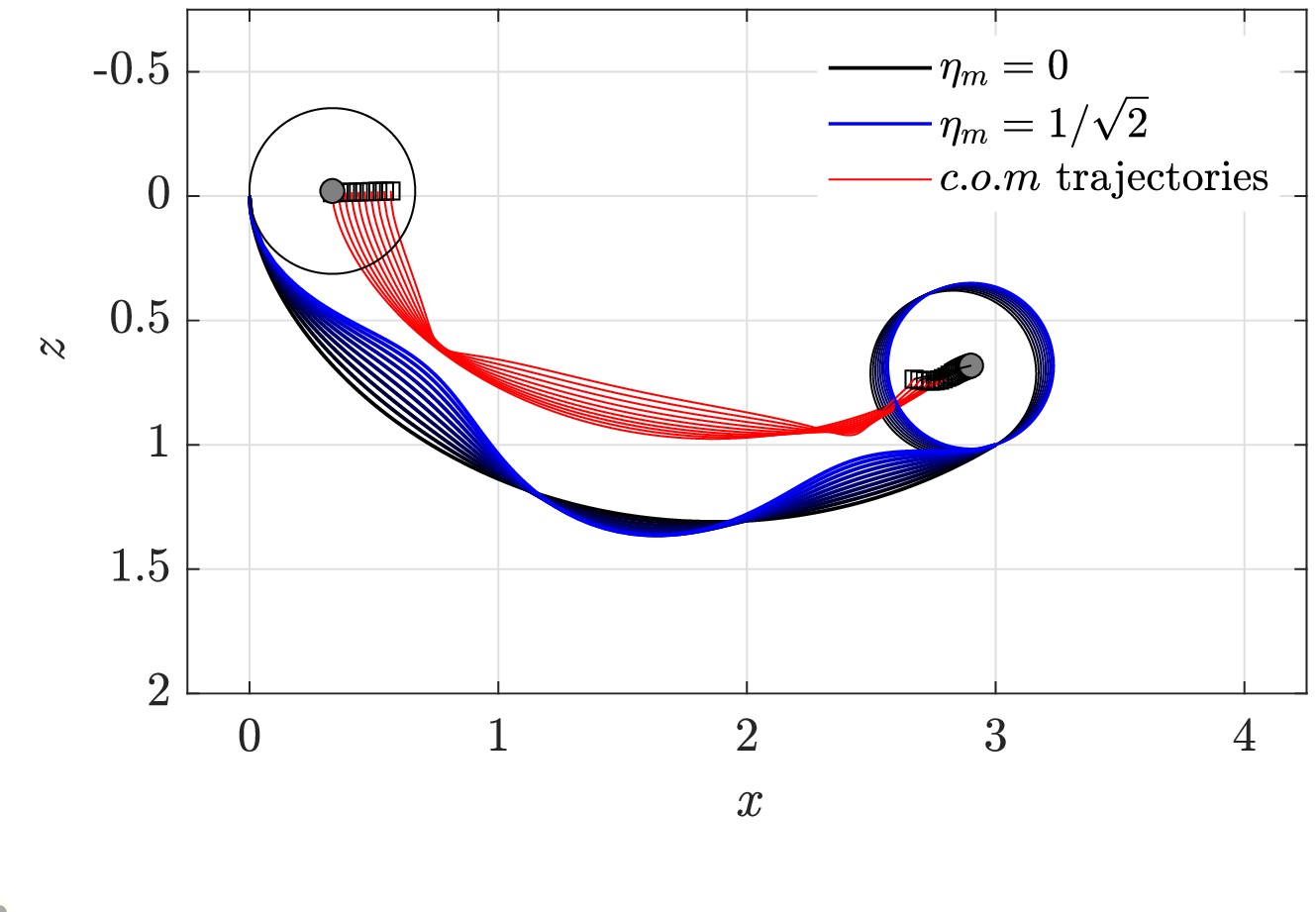}
    \caption{Brachistochrone paths and \textit{c.o.m} trajectories for the system of $L = 3$, $r = 1/3$, $\theta_m = 0$, $k = 0.5$ with $0\le \eta_m \le 1/\sqrt{2}$}
    \label{figEtaM}
\end{figure}

\reffig{figEtaM} shows brachistochrone paths for a family of cylinders when radial \textit{c.o.m} $\eta_m$ is varied from 0 (axis-symmetric configurations) to $\sqrt{1-k}$ (lumps of mass distributed along the rim of a mass-less rigid shell), with all other parameters held constant ($L=3$, $r=1/3$, $\theta_m = 0$, $k = 0.5$). This result illustrates the effect of off-centered \textit{c.o.m} in gradually inducing the peaks and valleys into its brachistochrone, and making them more pronounced as $\eta_m$ increases.

\subsection{Effect of initial angular location of \textit{c.o.m}}

\begin{figure}
    \centering
    \includegraphics[width=0.48\textwidth]{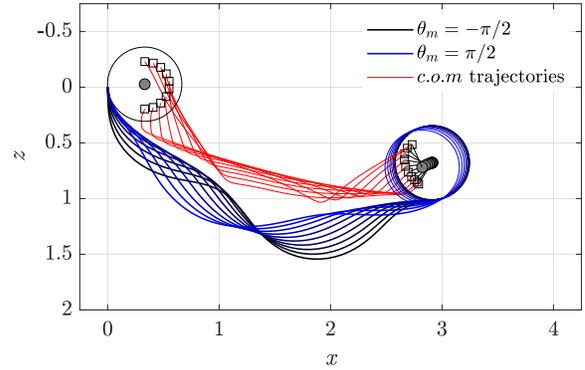}
    \caption{Brachistochrone paths and \textit{c.o.m} trajectories for the system of $L = 3$, $r = 1/3$, $\eta_m = 2/3$, $k = 0.5$ with $-\pi/2 \le \theta_m \le \pi/2$}
    \label{figThetaM}
\end{figure}

\reffig{figThetaM} shows brachistochrone paths for a family of off-centered cylinders when initial location of angular \textit{c.o.m} $\theta_m$ varies from $-\pi/2$ (\textit{c.o.m} is above $z=0$) to $\pi/2$ (\textit{c.o.m} is below $z=0$), with all other parameters held constant ($\eta_m = 2/3$, $r=1/3$, $L=3$, $k = 0.5$). While $\eta_m$ dictated the extent of peaks and valleys, this result shows that $\theta_m$ determines the location of those peaks and valleys.

\subsection{Effect of moment of inertia}

\begin{figure}
    \centering
    \includegraphics[width=0.48\textwidth]{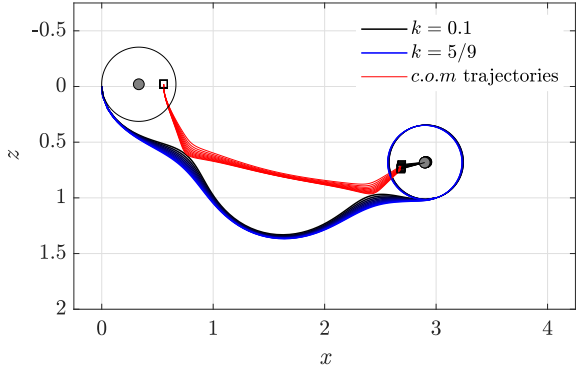}
    \caption{Brachistochrone paths and \textit{c.o.m} trajectories for the system of $L = 3$, $r = 1/3$, $\eta_m = 2/3$, $\theta_m = 0$ with $0.1 \le k \le 1 - \eta_m^2 = 5/9$}
    \label{figk}
\end{figure}

\reffig{figk} shows brachistochrone paths for a family of cylinders when $k$ is varied from 0.1 (mass mostly concentrated at $\eta_m$) to 1 - $\eta_m^2$ (lumps of mass distributed on the rim of a mass-less rigid shell), with all other parameters held constant ($\eta_m = 2/3$, $\theta_m = 0$, $r=1/3$, $L=3$). The  effect of $k$ on the brachistochrone is similar to that of $\eta_m$ but to a lesser degree.

\subsection{Effects of cylinder's radius}

\begin{figure}
 \centering
    \centering
    \includegraphics[width=0.48\textwidth]{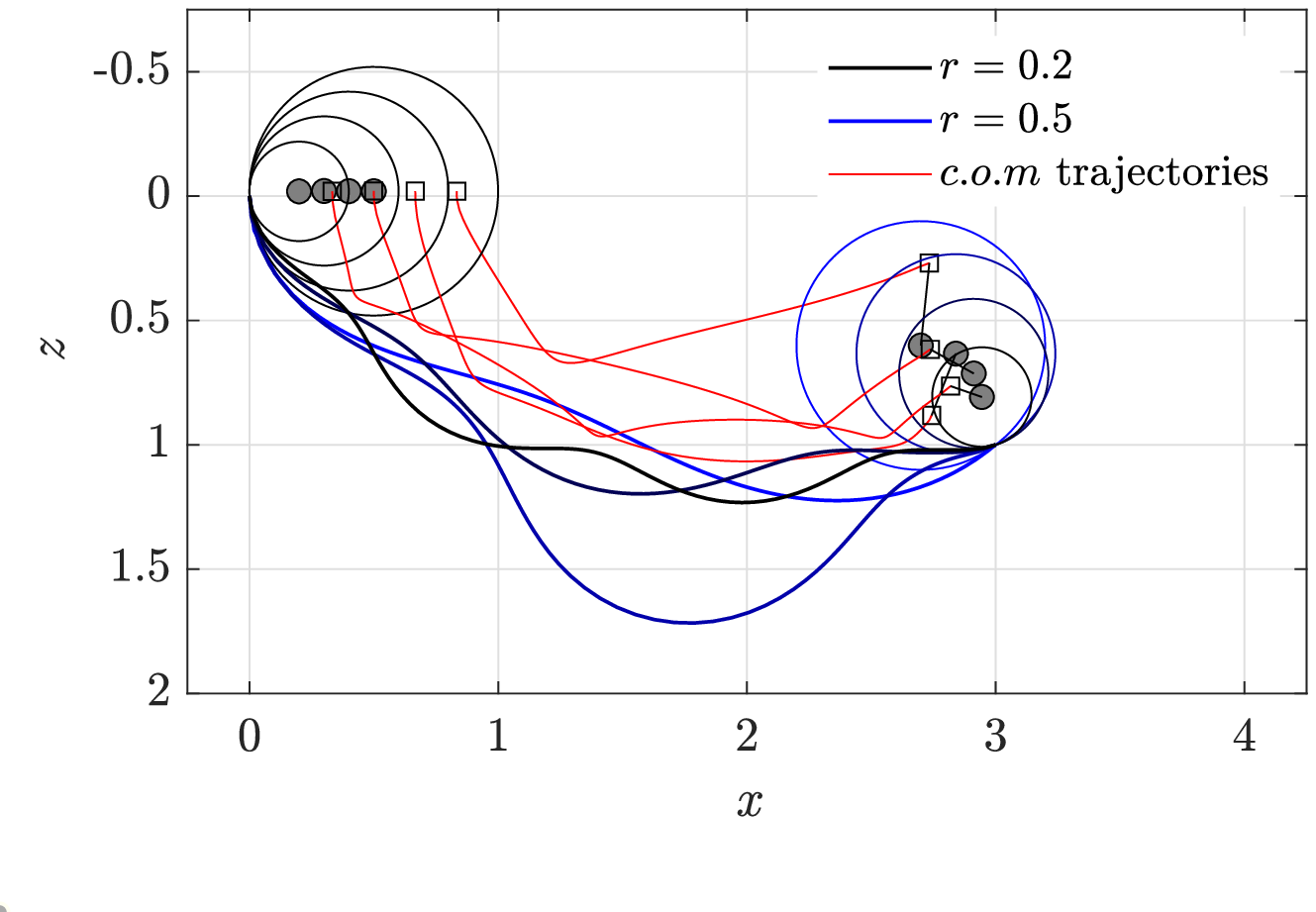}
    \caption{Brachistochrone paths and \textit{c.o.m} trajectories for the system of $L = 3$, $\eta_m = 2/3$, $\theta_m = 0$, $k = 0.5$ with $0.2\le r \le 0.5$}
    \label{figr}
\end{figure}

\reffig{figr} shows brachistochrone paths for a family of off-centered cylinders of increasing radii, $r$ from $0.2$ to $0.5$, with all other parameters held constant ($\eta_m = 1/3$  $\theta_m = 0$, $L=3$, $k = 0.5$). While $\eta_m, \theta_m$ and $k$ dictated the size and location of peaks and valleys in the brachistochrone, $r$ influences the $number$ of peaks and valleys with smaller cylinders generally having more of them. There is dual effect of $r$ on the solutions as it increases both the size of the cylinder and the radial \textit{c.o.m} $r_m = \eta_m r$.  The transit times are also seen to vary with $r$, unlike their axisymmetric counterparts whose transit times are independent of $r$\cite{legeza2010brachistochrone}.

\subsection{Effects of domain length}

\begin{figure}
    \centering
    \includegraphics[width=0.48\textwidth]{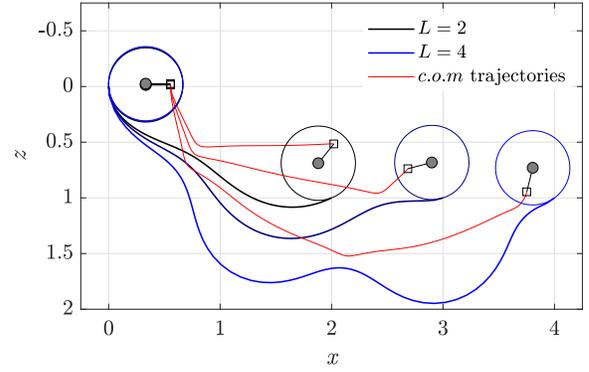}
    \caption{Brachistochrone paths and \textit{c.o.m} trajectories for the system of $r = 1/3$, $\eta_m = 2/3$, $\theta_m = 0$, $k = 0.5$ with $2\le L \le 4$}
    \label{figx2}
\end{figure}

\reffig{figx2} shows brachistochrone paths for an off-centered cylinder over different domain lengths ($L$), with all other parameters held constant ($\eta_m = 2/3$  $\theta_m = 0$, $r=1/3$, $k = 0.5$). As expected, longer domains result in longer paths for brachistochrones and longer transit times (3.96, 4.99 and 6.26 for $L = 2, 3$ and $4$ respectively).

\subsection{Effect of \textit{c.o.m} locations on shortest transit times}

\begin{figure}
    \centering
    \includegraphics[width=0.48\textwidth]{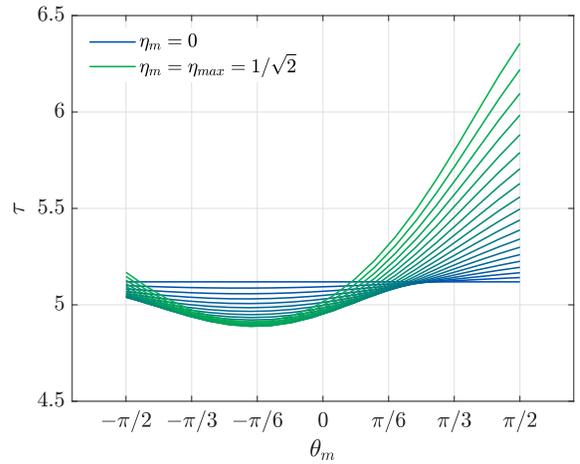}
    \caption{Variation of shortest transit times for off-centered cylinders ($r=1/3$, $k = 0.5$) with varying initial locations of \textit{c.o.m}. ($L=3$)}
    \label{figpp}
\end{figure}

Following specific case studies in prior subsections, a wider parametric space is explored in this subsection. The domain length $L$ and radius of the cylinder $r$ are fixed to 3 and 1/3 respectively. The moment of inertia $k$ is set to 0.5 (same as that for an uniform cylinder). The initial locations of \textit{c.o.m} in radial ($\eta_m$) and angular ($\theta_m$) directions are varied. For $\eta_m$, the physically allowable range is $[0, \sqrt{1-k}]$, considering all extremities of mass distribution. For $\theta_m$, the full range of $[-\pi,\pi]$ is physically allowable but there are other restrictions. When $\theta_m \in [-\pi/2,\pi/2]$, initial motion of both \textit{c.o.c} and \textit{c.o.m} with respect to \textit{c.o.c} is downwards. This ensures that the net initial motion of \textit{c.o.m} is downwards, in the direction of gravity. When $\theta_m \in [-\pi, -\pi/2) \cup (\pi/2, \pi]$, initial motion of \textit{c.o.c} would still be downwards but \textit{c.o.m} would be moving upwards with respect to \textit{c.o.c}. Depending on the value of $\eta_m$ and the path taken, these cases may result in aphysical scenarios (net upward initial motion of \textit{c.o.m}, against gravity) or violation of no-slip assumption, equations (\ref{eqnNoslip}) and (\ref{eqnThetaInt}). Therefore, parameter range of $\theta_m$ is restricted to $[-\pi/2, \pi/2]$ in this analysis.

For each pair of $(\eta_m, \theta_m)$ in the parametric space of $0 \le \eta_m \le 1/\sqrt{2}$ and $-\pi/2 \le \theta_m \le \pi/2$, brachistochrone paths are obtained and corresponding shortest transit times are computed. \reffig{figpp}  present a summary of shortest transit times. The first inference from this plot is that when the cylinder's \textit{c.o.m} coincides with its \textit{c.o.c} ($\eta_m = 0$), transit time $\tau$ is independent of initial angular location $\theta_m$ as expected (and equals that of an uniform cylinder, 5.12). Secondly, a more interesting result is that as $\eta_m$ increases, the effect of $\theta_m$ becomes more important. While lower $\theta_m$ indicates higher initial potential energy which can translate to higher kinetic energy and consequently shorter transit times, $\tau$ does not monotonically vary with $\theta_m$ but has a minimum. Thirdly, the location of this minimum does not vary with $\eta_m$ but remains constant at $\theta_m = -\pi/6$. These trends demonstrate that when a cylinder's mass is non-uniformly distributed, its initial orientation plays a critical role in determining its brachistochrone path and corresponding shortest transit time.

While brachistochrone transit times of axisymmetric objects are known to be always higher than that of a bead\cite{rodgers1946brachistochrone}, the same need not hold true for off-centered circular objects. For example, for the case of $k = 0.1$, $\eta_m = 4/5$, $\theta_m=-\pi/6$, $r = 1/3$, the off-centered cylinder has transit time of $\tau = 4.06$ which is quicker than that of bead (4.52) over the same domain ($L = 3$). 

\section{Conclusion}

Brachistochrones of off-centered cylinders, which hitherto have been an unopened and untapped system, present a treasure trove of problems rich in non-linear rigid body dynamics. Trajectories and transit times of off-centered cylinders have been shown to differ significantly from their axisymmetric counterparts. Systematic exploration of this multi-parameter problem demonstrated the sensitivity of its solutions to those parameters, in particular to that of \textit{c.o.m} location. Behavior of such systems in more complex force fields\cite{parnovsky1998some} prove to be challenging extensions. Off-centered cylinders also raise other interesting questions such as `what is the quickest brachistochrone ever achievable by \textit{any} object between two given points?' The method presented in section \ref{sPF} is general and extensible to tackle a spectrum of problems in this genre.

\begin{acknowledgments}
First author (KS) gratefully thanks his mentors Mr. Ethiraj Muthumalla (Retired Deputy Engineering Manager, Germanischer Llyod) for grooming KS' interest in mathematics, Prof. T. R. Subramaniam (Director of TRS IIT Classes) with whom KS had the privilege of learning fundamentals of calculus and Prof. Osman Basaran (Purdue University) who played a pivotal role in KS learning numerical methods and the arc length formulation, all of which have been instrumental in bringing this work to fruition.
\end{acknowledgments}

\appendix

\section{Types of off-centered cylinders}

\begin{figure}
\centering
\includegraphics[width=0.48\textwidth]{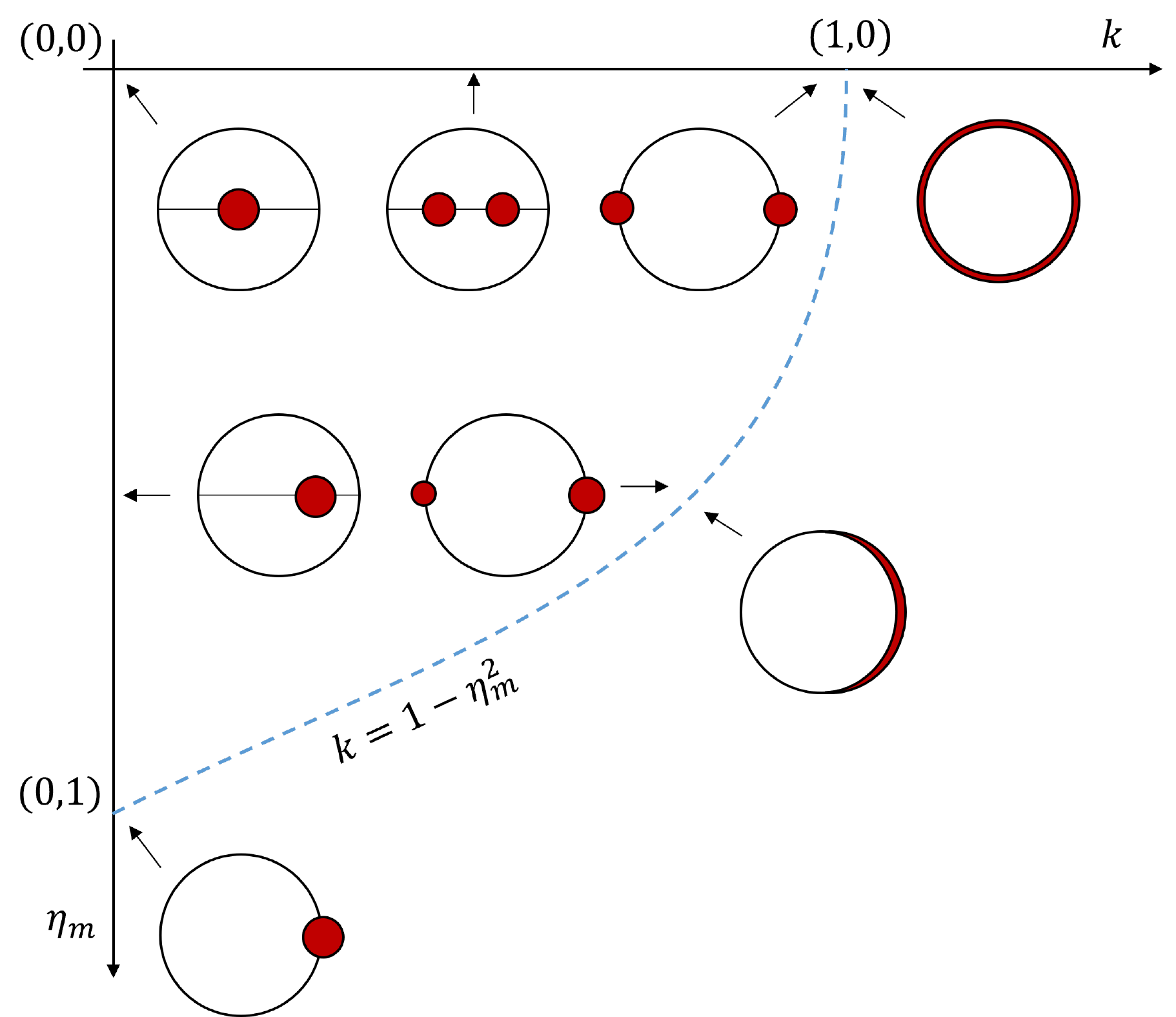}
\caption{Example configurations of cylinders, all with same mass $m$ and radius $r$, but varying moments of inertia ($k$) and center of mass locations ($\eta_m$)}
\label{figOCtypes}
\end{figure}

Example configurations of cylinders, all with same mass $m$ and radius $r$, but varying moments of inertia ($k$) and center of mass locations ($\eta_m$) are illustrated in \reffig{figOCtypes}. The red dots/circles represent concentrated masses (enlarged for easier visualization) and black circle/lines represent massless frames. There can exist multiple configurations of off-centered cylinders for the same value of $\eta_m$ and $k$ (as shown for $k=1$, $\eta_m = 0$ in the figure). Outside of the domain enveloped by blue dashed line ($\eta_m=0$, $k=0$ and $k=1-\eta_m^2$), no real configurations of off-centered cylinders can exist.

\section{Derivation of \refeqn{eqnTransitTime}}

\noindent
Differentiating equations (\ref{eqnxcm}) and (\ref{eqnzcm}) with time gives

\begin{align}
    \dot{x}_{cm} & = \dot{x}_c - r_m \dot{\theta} \sin \theta \notag \\
    \dot{z}_{cm} & = \dot{z}_c + r_m \dot{\theta} \cos \theta \notag
\end{align}

\noindent
Plugging the above expressions in \refeqn{eqnvcm} gives

\begin{align}
    v_{cm}^2 & = \dot{x}_c^2+r_m^2\dot{\theta}^2\sin^2\theta - 2 r_m \dot{\theta} \dot{x}_c\sin\theta \notag\\
    & \hspace{10pt} + \dot{z}_c^2+r_m^2\dot{\theta}^2\cos^2\theta + 2 r_m \dot{\theta} \dot{z}_c\cos\theta \notag \\
    & = v_c^2 + r_m^2 \dot{\theta}^2 - 2r_m\dot{\theta}(\dot{x}_c\sin \theta - \dot{z}_c \cos \theta) \label{eqnvcm2}
\end{align}

\noindent
The relationship that slope of the curve $z_c' = \dot{z}_c/\dot{x}_c$ and \refeqn{eqnv} for $v_c$ yield following equations for $\dot{x}_c$ and $\dot{z}_c$.

\begin{align}
    \dot{x}_c & = v_c/\sqrt{1+z_c'^2} \notag \\
    \dot{z}_c & = v_cz_c'/\sqrt{1+z_c'^2} \notag
\end{align}

\noindent
Plugging the above expressions along with \refeqn{eqnNoslip} for $\dot{\theta}$ in \refeqn{eqnvcm2} and replacing $r_m/r$ with $\eta_m$ gives

\begin{equation}
    v_{cm}^2 = v_c^2+\eta_m^2v_c^2 - 2\eta_mv_c^2(\sin\theta-z_c'\cos\theta)/\sqrt{1+z_c'^2} \notag
\end{equation}

\noindent
Defining $\beta$ using \refeqn{eqnbeta} simplifies the above equation to

\begin{equation}
    v_{cm}^2 = v_c^2(1+\eta_m^2 - 2\eta_m\sin(\theta-\beta)) \notag
\end{equation}

\noindent
Plugging the above expression in \refeqn{eqnEnergyBalance} gives

\begin{align}
    \frac{1}{2}mv_c^2(1+\eta_m^2 - 2\eta_m\sin(\theta-\beta)) + \frac{1}{2}kmr^2\omega^2 \notag\\
    = mg(z_c+r_m(\sin\theta-\sin\theta_m)) \notag
\end{align}

\noindent
Using \refeqn{eqnNoslip} for $\omega$ and rewriting the above equation for $v_c$ gives
\begin{equation}
    v_c = \left(\frac{2g(z_c+r_m(\sin\theta-\sin\theta_m))}{1+\eta_m^2 - 2\eta_m\sin(\theta-\beta) + k}\right)^{1/2} \label{eqnvc2}
\end{equation}

\noindent
The total time of transit for the cylinder between the two points is then given by
\begin{equation}
    T = \int_0^S\frac{ds_c}{v_c} \notag
\end{equation}
where $ds_c$ represents the incremental distance traversed by the \textit{c.o.c}. Using $ds_c = \sqrt{1+z_c'^2}dx_c$ and \refeqn{eqnvc2} for $v_c$, we obtain \refeqn{eqnTransitTime}.

\bibliography{references}

\end{document}